# THE TASK OF THE RELATIVISTIC OSCILLATOR IN A NON-INERTIAL FRAME OF REFERENCE

Perepelkin E. E., Sadovnikov B. I., Inozemtseva N. G.

**Annotation.** In this paper the analogues of the Lorentz transformations for non-inertial reference frames have been obtained. A common case when the movement speed of one coordinate frame in relation to another one can have time derivatives of higher orders. The obtained transformations conserve invariance of the space-time interval, and in the particular case of inertial frames become the well-known Lorentz transformations.

It is shown that the transition from classical mechanics to the theory of relativity is analogous to the consideration of the vibrating system described by the equation of the sine-Gordon type. In this case, if the amplitude of the elliptic functions is $k \to 0$ the fluctuations can be considered small, and that leads to classical mechanics. With $k \to 1$ time depends on the vibration amplitude, which leads to the theory of relativity. In the case of inertial frames the amplitude is $k = \beta = v/c$.

**Introduction.**

Construction of an adequate mathematical apparatus is significant while considering processes characterized by acceleration and accelerations of higher orders, such as the power of synchrotron radiation $n \sim \dot{v}^2$ [1-6]. In such cases it makes sense to consider not the classical phase space of coordinates and velocity (momentum), but the generalized phase space [7,8], the elements of which are $\vec{\xi} = \{\vec{r}, \vec{v}, \dot{\vec{v}}, ...\}^T$. In this space the generalized Liouville's theorem on conservation of the generalized phase volume has been applied. In the generalized phase space-time elements $(\vec{\xi}, ct)$ will have a non-linear law of transformation at the transition from one coordinate system to another [9]. The nonlinearity is caused by the presence of acceleration and accelerations of higher orders, which leads to consideration of non-inertial frames.

Consideration of non-inertial frames is usually performed in the framework of the General relativity [10-13]. However, the special relativity includes such possibility as well [14-16]. As it is known in the special relativity at the transition from one inertial coordinate frame $K_1$ to another inertial coordinate frame $K_2$ that is moving relatively $K_1$ with constant velocity $\vec{v}_0$ the Lorentz transformation is applied. In this paper we consider the case when the coordinate frame $K_1$ is inertial, and the coordinate frame $K_2$ is non-inertial in the general case, that is moving relatively $K_1$ with velocity $\vec{V}_{K_2}(t)$. The function $\vec{V}_{K_2}(t)$ will be considered analytic in time $t$, that is [9]:

$$\vec{V}_{K_2}(t) = \vec{v}_0 + \dot{\vec{v}}_0 t + \ddot{\vec{v}}_0 \frac{t^2}{2!} + \dddot{\vec{v}}_0 \frac{t^3}{3!} + ...,$$

where $\vec{v}_0, \dot{\vec{v}}_0, \ddot{\vec{v}}_0, \dddot{\vec{v}}_0, ...$ are constant values that determine the initial velocity, acceleration and acceleration of higher orders of the coordinate frame $K_2$ in relation to $K_1$. The analyticity of function $\vec{V}_{K_2}(t)$ allows one to mathematically rigorous consider not only the case of uniformly accelerated motion, but also the general case of accelerated motion, containing time derivatives of all orders.

As the Lorentz transformations are the generalization of the Galilean transformation, §1 deals with the generalized Galilean transformation for non-inertial frames.

§2 presents the analogue to the Lorentz transformations for non-inertial frames in differential form basing on the generalized Galilean transformations from §1. To derive these transformations the condition of invariance of the space-time interval is used. In the particular case of the inertial reference frame ($\dot{\vec{v}}_0 = \ddot{\vec{v}}_0 = ... = \vec{0}$) the received transformations become the famous Lorentz transformation. §2 also provides the conclusion of the transformation law of velocities in the case of non-inertial reference frame, which in the particular case of the inertial reference frame becomes the transformation law of velocities known in the special relativity. Similarly it is possible to obtain transformation laws of acceleration and the acceleration of higher orders.

§3 presents an example of the vibrating system, where velocities are compared with the speed of light, thus there is acceleration. The simplest example of such systems can be ring accelerators, or atomic structures [17]. It was found that time in such systems is connected with the amplitude of oscillations. So, while considering the oscillations described by the sine-Gordon equation [18-20], the period (time) depends on the amplitude of oscillation. The same situation occurs at speeds close to the speed of light, as a result, time in the accelerated system flows slower. At low amplitude of oscillations, the sine-Gordon equation becomes the linear equation, so the period (time) does not depend on the amplitude of oscillations. In this case, there is the classical Newton mechanics, and time in all systems flows equally. As the solution of the sine-Gordon equation are expressed through elliptic functions having amplitude $k$ as parameter [21-26], at the transition from non-relativism to relativism, amplitude $k$ changes from 0 to 1. A similar situation, as it is known, takes place for solutions of the sine-Gordon equation, when functions of elliptic sine and cosine are transferred into the corresponding trigonometric functions, and the pendulum can be considered as mathematical one. In the particular case of the inertial frame, with the Lorentz transformation laws the amplitude was $k = \beta = v/c$.

### § 1. Galilean transformations for non-inertial reference frames

In the case of inertial reference systems, the movement of the coordinate frame $K_2$ in relation to the coordinate frame $K_1$ is described by the classical Galilean transformation

$$\vec{R}_{K_2}(t) = \vec{r}_0 + \vec{v}_0 t, \tag{1}$$

where $\vec{r}_0, \vec{v}_0$ − constant values. Value $\vec{r}_0$ corresponds with the position of the origin of coordinate frame $K_2$ at the initial moment of time $t = 0$ in the coordinate frame $K_1$, that is $\vec{R}_{K_2}(0) = \vec{r}_0$. Value $\vec{v}_0$ sets the movement speed of the coordinate frame $K_2$ is relatively the coordinate frame $K_1$.

Let us consider the case where the coordinate frame moves along a smooth (analytical trajectory) is relation to the coordinate frame $K_1$ with speed $\vec{V}_{K_2}(t)$, then instead of (1) it can be

$$\vec{R}_{K_2}(t) = \vec{r}_0 + \vec{v}_0 t + \dot{\vec{v}}_0 \frac{t^2}{2} + \ddot{\vec{v}}_0 \frac{t^3}{3!} + ..., \tag{2}$$

where constant values $\dot{\vec{v}}_0, \ddot{\vec{v}}_0, ...$ correspond to the initial acceleration and the initial acceleration of a higher order. A particular case of expression (2) (when $\dot{\vec{v}}_0 = \ddot{\vec{v}}_0 = ... = \vec{0}$) is the expression (1), that corresponds to the classical Galilean transformation. As a result, the center of the coordinate frame $K_1$ moves along the trajectory (2).

Let the movement of a kinematic point $p$ in the coordinate frame $K_1$ is described by a smooth (analytic) trajectory $\vec{R}_p(t)$, and in the coordinate frame $K_2$ by the trajectory $\vec{r}_p(t)$:

$$\vec{R}_p(t) = \vec{R} + \vec{V}t + \dot{\vec{V}}\frac{t^2}{2} + ...,$$
$$\vec{r}_p(t) = \vec{r} + \vec{v}t + \dot{\vec{v}}\frac{t^2}{2} + ...,$$
(3)

where constant values $\vec{R}, \vec{V}, \dot{\vec{V}}, ...$ and $\vec{r}, \vec{v}, \dot{\vec{v}}, ...$ correspond to the initial velocity, initial acceleration and initial accelerations of higher order in the coordinate frames $K_1$ and $K_2$ respectively for the kinematic point $p$.

In the initial moment of time $t=0$ coordinates of the kinematic point $p$ in frames $K_1$ and $K_2$ have the meanings, respectively: $\vec{R}_p(0) = \vec{R}$ and $\vec{r}_p(0) = \vec{r}$. As the origin of the coordinate frame $K_2$ is moving relatively $K_1$ along the trajectory $\vec{R}_{K_2}$ (2), at the initial moment of time $t=0$, taking (3) into account, the following ratio is true

$$\vec{R} = \vec{r}_0 + \vec{r} \quad or \quad \vec{R}_p(0) = \vec{R}_{K_2}(0) + \vec{r}_p(0).$$
(4)

At a random moment of time expression (4) is the following:

$$\vec{R} + \vec{V}t + \dot{\vec{V}}\frac{t^2}{2} + ... = \vec{r}_0 + \vec{v}_0 t + \dot{\vec{v}}_0 \frac{t^2}{2} ... + \vec{r} + \vec{v}t + \dot{\vec{v}}\frac{t^2}{2} + ...$$
(5)

or

$$\vec{R}_p(t) = \vec{R}_{K_2}(t) + \vec{r}_p(t).$$

The validity of (5) directly follows from (4), if multiplying the equation from (4): $t^0, t, \frac{t^2}{2}, ..., \frac{t^n}{n!}, ...$, and then summarizing, then we obtain (5). Due to the analyticity of trajectories the series in (5) are convergent in some point of $t=0$ [7,9].

If we add the relation $T_p = t$ to the relations (5), we obtain the Galilean transformation for non-inertial frames:

$$\vec{R}_p = \vec{r}_p + \vec{r}_0 + \vec{v}_0 t + \dot{\vec{v}}_0 \frac{t^2}{2!} + \ddot{\vec{v}}_0 \frac{t^3}{3!} + ...,$$
$$T_p = t,$$

or, delete the index $p$ for convenience writing,

$$\vec{R} = \vec{r} + \vec{r}_0 + \vec{v}_0 t + \dot{\vec{v}}_0 \frac{t^2}{2!} + \ddot{\vec{v}}_0 \frac{t^3}{3!} + ..., \qquad (6)$$
$$T = t.$$

In the particular case of inertial reference frames ($\dot{\vec{v}}_0 = \ddot{\vec{v}}_0 = ... = \vec{0}$) transformation (6) becomes the classical Galilean transformation:

$$\vec{R} = \vec{r} + \vec{r}_0 + \vec{v}_0 t, \qquad (7)$$
$$T = t.$$

## § 2. The space-time transformations for non-inertial reference frames

Let us find transformations for non-inertial reference frames conserving the invariance of the space-time interval. In the coordinate frame $K_1$ the space-time interval has the form:

$$(ds)^2 = c^2 (dT)^2 - (d\vec{R})^2. \qquad (8)$$

Substituting the expressions of the coordinates into the space-time interval (8) instead of coordinates $\vec{R}$ through coordinates $\vec{r}$ using the generalized Galilean transformation (6) and obtaining the conditions under which the space-time interval will remain invariant.

$$(d\vec{R})^2 = (d\vec{r} + d\vec{R}_{K_2})^2 = (d\vec{r})^2 + 2(d\vec{r}, d\vec{R}_{K_2}) + (d\vec{R}_{K_2})^2. \qquad (9)$$

Consider that

$$d\vec{R}_{K_2} = \left(\vec{v}_0 + \dot{\vec{v}}_0 t + \ddot{\vec{v}}_0 \frac{t^2}{2} ...\right) dt = \vec{V}_{K_2}(t) dt. \qquad (10)$$

Substitute (10) into (9), and substitute the obtained result into (8).

$$(d\vec{R})^2 = (d\vec{r})^2 + 2(d\vec{r}, \vec{V}_{K_2}) dt + V_{K_2}^2 (dt)^2, \qquad (11)$$
$$(ds)^2 = c^2 (dt)^2 - (d\vec{r})^2 - 2(d\vec{r}, \vec{V}_{K_2}) dt - V_{K_2}^2 (dt)^2 =$$
$$= c^2 (dt)^2 \left(1 - \frac{V_{K_2}^2}{c^2}\right) - 2(d\vec{r}, \vec{V}_{K_2}) dt - (d\vec{r})^2. \qquad (12)$$

Next, identifying the full square in expression (12), and we obtain:

$$(ds)^2 = c^2 \left(dt \sqrt{1 - \frac{V_{K_2}^2}{c^2}} - \frac{(d\vec{r}, \vec{V}_{K_2})}{c^2 \sqrt{1 - \frac{V_{K_2}^2}{c^2}}}\right)^2 - \frac{(d\vec{r}, \vec{V}_{K_2})^2}{c^2 \left(1 - \frac{V_{K_2}^2}{c^2}\right)} - (d\vec{r})^2. \qquad (13)$$

Note that the space-time interval (8) or (13) consists of two parts: positive and negative one. The positive part corresponds to the time value and a negative – to coordinate space. Therefore, we introduce new variables:

$$dT' \stackrel{det}{=} dt\sqrt{1-\frac{V_{K_2}^2}{c^2}} - \frac{(d\vec{r},\vec{V}_{K_2})}{c^2\sqrt{1-\frac{V_{K_2}^2}{c^2}}}, \quad (d\vec{R}')^2 \stackrel{det}{=} \frac{(d\vec{r},\vec{V}_{K_2})^2}{c^2\left(1-\frac{V_{K_2}^2}{c^2}\right)} + (d\vec{r})^2. \tag{14}$$

Then the expression for the space-time interval (13) with (14) takes the form:

$$(ds)^2 = c^2(dT')^2 - (d\vec{R}')^2. \tag{15}$$

As a result, two consecutive transformations (6) and (14) leave the metric forminvariant. Let us substitute expressions (6) into (14), and we obtain:

$$dT' = dT\sqrt{1-\frac{V_{K_2}^2}{c^2}} - \frac{(d\vec{R},\vec{V}_{K_2}) - V_{K_2}^2 dT}{c^2\sqrt{1-\frac{V_{K_2}^2}{c^2}}}, \tag{16}$$

$$(d\vec{R}')^2 = \frac{(d\vec{R} - \vec{V}_{K_2}dT, \vec{V}_{K_2})^2}{c^2\left(1-\frac{V_{K_2}^2}{c^2}\right)} + (d\vec{R} - \vec{V}_{K_2}dT)^2. \tag{17}$$

Simplifying expressions (16) and (17). Let us bring expression (16) to a common denominator, and we get:

$$dT' = \frac{c^2\left(dT\left(1-\frac{V_{K_2}^2}{c^2}\right) - \left(d\vec{R},\frac{\vec{V}_{K_2}}{c^2}\right) + \frac{V_{K_2}^2}{c^2}dT\right)}{c^2\sqrt{1-\frac{V_{K_2}^2}{c^2}}} = \frac{dT - \left(d\vec{R},\frac{\vec{V}_{K_2}}{c^2}\right)}{\sqrt{1-\frac{V_{K_2}^2}{c^2}}}. \tag{18}$$

Let us introduce the designation:

$$\gamma_0(T) \stackrel{det}{=} \frac{1}{\sqrt{1-\frac{V_{K_2}^2(T)}{c^2}}}. \tag{19}$$

Then (18) takes the form:

$$dT' = \gamma_0(T)\left(dT - \left(d\vec{R},\frac{\vec{V}_{K_2}(T)}{c^2}\right)\right). \tag{20}$$

Simplifying expression (17):

$$\left(d\vec{R}-\vec{V}_{K_2}dT\right)^2 + \frac{\gamma_0^2}{c^2}\left(d\vec{R}-\vec{V}_{K_2}dT,\vec{V}_{K_2}\right)^2 = \left(d\vec{R}\right)^2 - 2\left(d\vec{R},\vec{V}_{K_2}\right)dT + V_{K_2}^2(dT)^2 +$$

$$+\frac{\gamma_0^2}{c^2}\left[\left(d\vec{R},\vec{V}_{K_2}\right)^2 - 2\left(d\vec{R},\vec{V}_{K_2}\right)V_{K_2}^2 dT + V_{K_2}^4(dT)^2\right] = \left(d\vec{R}\right)^2 - 2\left(d\vec{R},\vec{V}_{K_2}\right)dT\left(1+\frac{\gamma_0^2}{c^2}V_{K_2}^2\right) + \quad (21)$$

$$+(dT)^2 V_{K_2}^2\left(1+V_{K_2}^2\frac{\gamma_0^2}{c^2}\right) + \frac{\gamma_0^2}{c^2}\left(d\vec{R},\vec{V}_{K_2}\right)^2 =$$

$$= \left(d\vec{R}\right)^2 + \frac{\gamma_0^2}{c^2}\left(d\vec{R},\vec{V}_{K_2}\right)^2 - 2\gamma_0^2\left(d\vec{R},\vec{V}_{K_2}\right)dT + (dT)^2 V_{K_2}^2\gamma_0^2.$$

Note that the first two summands in expression (21) can be presented in the following way:

$$\left[d\vec{R}+(\gamma_0-1)\frac{\vec{V}_{K_2}}{V_{K_2}^2}\left(d\vec{R},\vec{V}_{K_2}\right)\right]^2 = \left(d\vec{R}\right)^2 + 2(\gamma_0-1)\frac{\left(d\vec{R},\vec{V}_{K_2}\right)^2}{V_{K_2}^2} +$$

$$+(\gamma_0-1)^2\frac{V_{K_2}^2}{V_{K_2}^4}\left(d\vec{R},\vec{V}_{K_2}\right)^2 = \left(d\vec{R}\right)^2 + \frac{\gamma_0^2}{c^2}\left(d\vec{R},\vec{V}_{K_2}\right)^2. \quad (22)$$

Therefore, expression (17) is rearranged in the form:

$$\left(d\vec{R}'\right)^2 = \left(d\vec{R}-\vec{V}_{K_2}dT\right)^2 + \frac{\gamma_0^2}{c^2}\left(d\vec{R}-\vec{V}_{K_2}dT,\vec{V}_{K_2}\right)^2 =$$

$$= \left[d\vec{R}+(\gamma_0-1)\frac{\vec{V}_{K_2}}{V_{K_2}^2}\left(d\vec{R},\vec{V}_{K_2}\right)\right]^2 - 2\gamma_0^2\left(d\vec{R},\vec{V}_{K_2}\right)dT + (dT)^2 V_{K_2}^2\gamma_0^2 = \quad (23)$$

$$= \left[d\vec{R}+(\gamma_0-1)\frac{\vec{V}_{K_2}}{V_{K_2}^2}\left(d\vec{R},\vec{V}_{K_2}\right) - \vec{V}_{K_2}\gamma_0 dT\right]^2.$$

As a result, we obtain the expression:

$$d\vec{R}' = d\vec{R}+(\gamma_0(T)-1)\frac{\vec{V}_{K_2}(T)}{V_{K_2}^2(T)}\left(d\vec{R},\vec{V}_{K_2}(T)\right) - \vec{V}_{K_2}(T)\gamma_0(T)dT. \quad (24)$$

***Let us name expressions (20) and (24) as a differential form of the space-time transformations for non-inertial frames.*** Note that the obtained transformations (20), (24) move into the classical Lorentz transformations in case of the inertial frames consideration ( $\dot{\vec{v}}_0 = \ddot{\vec{v}}_0 = ... = \vec{0}$ ). Indeed

$$\vec{V}_{K_2}(T) = \vec{v}_0, \quad \gamma_0(T) = \frac{1}{\sqrt{1-\frac{V_{K_2}^2(T)}{c^2}}} = \frac{1}{\sqrt{1-\frac{v_0^2}{c^2}}} = \gamma, \quad (25)$$

Substituting (25) into (20) and (24), we obtain

$$dT' = \gamma\left(dT - \left(d\vec{R}, \frac{\vec{v}_0}{c^2}\right)\right),$$
$$d\vec{R}' = d\vec{R} + (\gamma-1)\frac{\vec{v}_0}{v_0^2}(d\vec{R}, \vec{v}_0) - \vec{v}_0 \gamma dT. \tag{26}$$

Making the integration in (26), we obtain the final form of the classic Lorentz transformations for inertial frames:

$$T' = \gamma\left(T - \left(\vec{R}, \frac{\vec{v}_0}{c^2}\right)\right),$$
$$\vec{R}' = \vec{R} + (\gamma-1)\frac{\vec{v}_0}{v_0^2}(\vec{R}, \vec{v}_0) - \vec{v}_0 \gamma T. \tag{27}$$

Obtaining the ratio for converting of speeds.

$$\vec{V}' \stackrel{\text{det}}{=} \frac{d\vec{R}'}{dT'} = \vec{V}\frac{dT}{dT'} + (\gamma_0(T)-1)\frac{\vec{V}_{K_2}(T)}{V_{K_2}^2(T)}(\vec{V}, \vec{V}_{K_2}(T))\frac{dT}{dT'} - \vec{V}_{K_2}(T)\gamma_0(T)\frac{dT}{dT'}, \tag{28}$$

taking into account

$$\frac{dT'}{dT} = \gamma_0(T)\left(1 - \left(\vec{V}, \frac{\vec{V}_{K_2}(T)}{c^2}\right)\right),$$

we obtain

$$\vec{V}' = \frac{\vec{V} + (\gamma_0(T)-1)\dfrac{\vec{V}_{K_2}(T)}{V_{K_2}^2(T)}(\vec{V}, \vec{V}_{K_2}(T)) - \vec{V}_{K_2}(T)\gamma_0(T)}{\gamma_0(T)\left(1 - \dfrac{(\vec{V}, \vec{V}_{K_2}(T))}{c^2}\right)}. \tag{29}$$

Expression (29) defines the velocities transformation law for non-inertial frames. A similar method can be used for obtaining the transformation for acceleration and for acceleration of higher orders.

Note that expression (29) in the case of inertial frames ($\dot{\vec{v}}_0 = \ddot{\vec{v}}_0 = ... = \vec{0}$) passes to the well-known velocities transformation law in the special relativity

$$\vec{V}' = \frac{\vec{V} + (\gamma-1)\dfrac{\vec{v}_0}{v_0^2}(\vec{V}, \vec{v}_0) - \vec{v}_0 \gamma}{\gamma\left(1 - \dfrac{(\vec{V}, \vec{v}_0)}{c^2}\right)}. \tag{30}$$

### §3 Example of the vibrating system

As we have obtained the transformation of coordinates for the arbitrary accelerated motion (20), (24), we can consider a famous the twin paradox [27-32] without any assumptions related to uniformly accelerated motion, or instantaneous velocity. Let the frame $K_1$ be inertial, and the frame $K_2$ moves relatively $K_1$ under the law

$$\vec{V}_{K_2}(T) = \{V_{max}\sin(\omega T), 0, 0\}, \qquad (31)$$

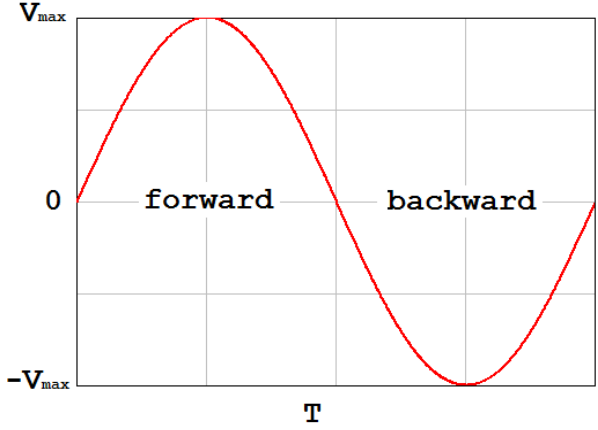

Fig. 1 Reference frame $K_2$ velocity

where the constant $V_{max}$ defines the maximum possible speed of motion of the frame $K_2$ relatively the frame $K_1$. The following condition is supposed to be fulfilled

$$k \stackrel{det}{=} \frac{V_{max}}{c} < 1. \qquad (32)$$

A constant value $\omega$ specifies the frequency of the oscillatory motion (31). As a result, at the initial moment of time the coordinate frame $K_2$ according to (31) rests relatively the frame $K_1$ (see Fig.1). At $0 < T < \frac{\pi}{2\omega}$ the frame $K_2$ moves with acceleration, and in the moment of time $T = \frac{\pi}{2\omega}$ reaches the maximum velocity $V_{max}$. At $\frac{\pi}{2\omega} < T < \frac{\pi}{\omega}$ movement of the frame $K_2$ becomes slower and at the moment of time $T = \frac{\pi}{\omega}$ the velocity is equal to zero and the frame $K_2$ stops. In the period $\frac{\pi}{\omega} < T < \frac{3\pi}{2\omega}$ the frame $K_2$ moves in the opposite direction – back to the frame $K_1$ with acceleration. At the moment of time $T = \frac{3\pi}{2\omega}$ the acceleration is equal to zero and the frame $K_2$ reaches the maximum velocity $-V_{max}$. Further, at $\frac{3\pi}{2\omega} < T < \frac{2\pi}{\omega}$ the motion of the frame $K_2$ slows down, and at the moment of time $T = \frac{2\pi}{\omega}$ the frame $K_2$ stops and returns to its original spatial position corresponding to the initial moment of time $T = 0$.

In the initial moment of time clocks in the frames $K_1$ and $K_2$ are synchronized. Let us rate the relative time difference $\eta = 1 - \frac{\Delta T}{\Delta T'}$ after returning of the coordinate frame $K_2$ to its original state. For simplicity of calculations, we assume that at the initial moment of time the coordinate origin of the frame $K_1$ and $K_2$ coincide, and clocks are located at the origin of coordinates.

Substitute the expression for the velocity (31) into the transformation (20), and perform the integration, we obtain:

$$dT' = \frac{dT}{\sqrt{1 - \frac{V_{max}^2}{c^2}\sin^2(\omega T)}} = \frac{1}{\omega}\frac{d\xi}{\sqrt{1 - k^2\sin^2\xi}},$$

$$\varphi' = \int_0^{\varphi'} d\xi' = \int_0^{\varphi} \frac{d\xi}{\sqrt{1-k^2 \sin^2 \xi}} = F(\varphi, k), \qquad (33)$$

$$\omega T' = F\left(\omega T, \frac{V_{max}}{c}\right), \quad \omega T = \operatorname{am}\left(\omega T', \frac{V_{max}}{c}\right), \qquad (34)$$

where $\xi = \omega T$, $\xi' = \omega T'$ are indicated. Function $F(\varphi, k)$ is an incomplete elliptic integral of the 1st kind, if there is condition (32). Thus, the time $T$ and $T'$ are connected through the relation (34). Using (34) we can rewrite a function of velocity (31) with respect to time $T'$

$$V_{K_2}^{(x)} = V_{max} \sin(\omega T) = V_{max} \sin(\operatorname{am}(\omega T', k)) = V_{max} \operatorname{sn}(\omega T'), \qquad (35)$$

or

$$\sin(\omega T) = \operatorname{sn}(\omega T'), \qquad (36)$$

where function sn is the elliptic sine. To calculate the total time, it is necessary to integrate into (33) from $0$ to $2\pi$. Basing on the form of the subintegral function in (33) we can integrate only from $0$ to $\frac{\pi}{2}$ and multiply the result by 4, i.e.

$$\omega \Delta T' = \int_0^{2\pi} \frac{d\xi}{\sqrt{1-k^2 \sin^2 \xi}} = 4 \int_0^{\pi/2} \frac{d\xi}{\sqrt{1-k^2 \sin^2 \xi}} = 4K(k), \qquad (37)$$

$$\Delta T' = \frac{4}{\omega} K(k), \quad \Delta T = \frac{2\pi}{\omega}, \qquad (38)$$

$$\eta = 1 - \frac{\pi}{2K(k)}, \qquad (39)$$

where $K(k)$ is the complete elliptic integral of the 1st kind. Note that expression (38) corresponds to expression (36), that $\sin(2\pi) = \operatorname{sn}(4K(k)) = 0$.

Fig. 2 shows the relative time dilation (39) in the coordinate frame $K_2$ with respect to time in the coordinate frame $K_1$ in percentage. It is seen that the time dilation is connected with coefficient $k$ (32). At $k=1$ the maximum velocity $V_{max} = c$ and the complete elliptic integral (37) diverges $K(1) = \infty$, that is $\Delta T' \to +\infty$. At $k=0$ the coordinate frame $K_2$ at rest with respect to the frame $K_1$ and $K(0) = \frac{\pi}{2}$ i.e. $\Delta T' = \Delta T$, therefore, the time difference is not observed.

Let us consider the result from relations (36), (38). The function $\operatorname{sn}(u)$ has the period $4K$ i.e. $\operatorname{sn}(u+4K) = \operatorname{sn}(u)$, so we choose $T' = T_0'$ in (36) so that $\omega T_0' = 4K(k)$, that is

$$T_0' = 4K(k)\frac{1}{\omega}. \qquad (40)$$

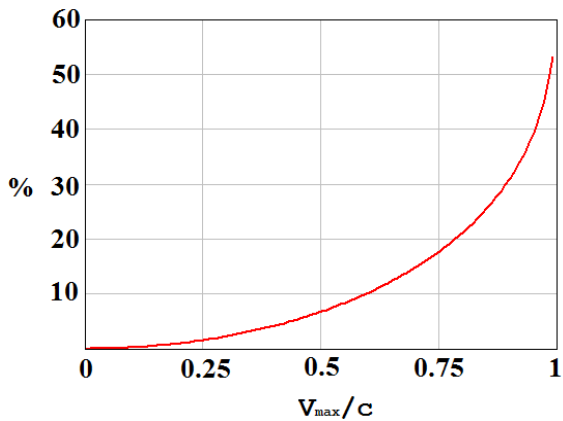

Fig.2 The relative time dilation

If we consider the equation of the pendulum oscillations without approximation of small oscillations (the sine-Gordon equation), then formula (40) determines the period of oscillation with the frequency $\omega = \sqrt{\dfrac{g}{l}}$, where $g$ is the gravitational acceleration and $l$ is the length of the pendulum. As a result, the period of oscillation in contrast to the mathematical pendulum according to (40) depends on the amplitude of oscillations $T'_0 = \dfrac{4K(k)}{\omega}$. At low amplitude of oscillations $k \to 0$, as mentioned above $K(k) \to \dfrac{\pi}{2}$ and the period of oscillations (40) turns into the classical period $T'_0 = \dfrac{2\pi}{\omega}$, independent from the amplitude of oscillations. As a result, when ( $V_{max} \ll c$ ($k \to 0$)) it turns out that the frame $K_2$ makes «small amplitude oscillations» ($k \to 0$) relatively the frame $K_1$ under the periodic law (31). In this case, the period (time) can be considered independent from the amplitude $k$, i.e. to assume that $\Delta T' = \Delta T$ in (38). At the oscillations of the «large amplitude» ($k \to 1$), that is, when $V_{max} \to c$ the period (time) essentially depends on the amplitude $k$, and $\Delta T' \neq \Delta T$ in (38). By the way, in the theory of elliptic functions, the value $k$ is called the amplitude. Thus, from the point of view of the theory of relativity, the amplitude $k$ defines the boundary between the classical Newton mechanics ($k \to 0$) and the theory of relativity ($k \to 1$). If we consider the Lorentz transformation with a constant velocity $V_{K_2} = Const$, then the amplitude is $k = \dfrac{V_{K_2}}{c} = \beta$.

In conclusion, let us consider another example illustrating the rotational movement. Let the coordinate frame $K_2$ move relatively the coordinate frame $K_1$ at a velocity of

$$\vec{V}_{K_2}(T) = \{V_{max}\sin(\omega T), V_{max}\cos(\omega T), 0\}. \qquad (41)$$

As before, we assume that at the initial moment of time the coordinate frames are converged. Movement (41) takes place in accelerator physics, for example, at consideration of a charged particle beam motion in the accelerator ring. In contrast to the previous case, the module of the velocity $|\vec{V}_{K_2}(T)| = V_{max} = Const$ does not depend on time. As in the previous case the similar (35) ratio is true

$$\vec{V}_{K_2}(T') = \{V_{max}\operatorname{sn}(\omega T'), V_{max}\operatorname{cn}(\omega T'), 0\}, \qquad (42)$$

note that the module of the velocity (42) remains unchanged, as $\operatorname{sn}^2(\omega T') + \operatorname{cn}^2(\omega T') = 1$. As in the previous case with $k = 0$ elliptic functions transform into trigonometric $\operatorname{sn}(\omega T', k=0) = \sin(\omega T')$, $\operatorname{cn}(\omega T', k=0) = \cos(\omega T')$, and $\Delta T' = \Delta T$. In the case of $k \to 1$ there is a relative change in time (39).